# Mode coupling theory analysis of water dynamics detected by dielectric spectroscopy

**P. Lunkenheimer**[a] **and A. Loidl**

**AFFILIATION**

Experimental Physics V, Center for Electronic Correlations and Magnetism, University of Augsburg, 86135 Augsburg, Germany

[a]**Author to whom correspondence should be addressed:** peter.lunkenheimer@physik.uni-augsburg.de

**ABSTRACT**

Mode-coupling theory (MCT), one of the most prominent models of the glass transition, also makes various specific predictions for the molecular dynamics in the liquid state of matter. They can be best experimentally tested by evaluating susceptibility spectra covering a broad frequency range. In the present work, we clarify whether the most recent broadband dielectric-spectroscopy results on water, whose glass transition is highly disputed, are compatible with MCT. The so-called fast $\beta$ process, expected within this theory, should be detectable at low temperatures only, inaccessible in water due to its inevitable crystallization upon cooling. Therefore, we also investigate data on aqueous LiCl solutions, which can be easily supercooled. We find that all the available broadband dielectric data on these solutions and pure water are well consistent with MCT. Notably, this includes the expected interrelation of the critical temperature dependence of the $\alpha$ relaxation to its often-neglected non-Debye spectral shape. Moreover, the available data on the fast $\beta$ process, detected in a 14.8 mol% LiCl solution, are also well consistent with the theory. We find a smooth development of the derived MCT parameters with salt concentration which supports the validity of the critical temperature $T_c = 222$ K and system parameter $\lambda = 0.55$ deduced here for pure water. We also compare the current dielectric relaxation-time results to those determined by other experimental methods and address the question whether information on the controversial fragile-to-strong transition of water can be obtained from such data. Finally, we discuss the significance of the dominant dielectric relaxation process of water for the glass transition as described by MCT, especially considering scenarios questioning its association with the $\alpha$ relaxation.

## I. INTRODUCTION

Among various competing models aiming at a better understanding of the glass transition and the glassy state of matter, mode-coupling theory (MCT)[1,2] has emerged as one of the most influential approaches.[3,4,5] It predicts a dynamical transition at a critical temperature $T_c$, located above the glass-transition temperature $T_g$, at which the actual crossover from a supercooled liquid to a glass (i.e., a solid without crystalline order) is experimentally observed. Within idealized MCT, $T_c$ marks a kind of "ideal" dynamic glass transition which should lead to total arrest of the molecule dynamics. However, in extended versions of the theory, this transition is assumed to be smeared out due to additional hopping processes that come into effect close to and below $T_c$.[2,6] This is in accord with experimental observations, which do not reveal any clear-cut anomalies at $T_c$ in temperature-dependent dynamic quantities like viscosity or relaxation time. While still controversial in many respects, nowadays some consensus seems to emerge that MCT is at least valid at high temperatures, in the low-viscosity liquid state, i.e., above $T_c$.[3,4,5] Thus, especially idealized MCT may be considered to be essentially a theory of the liquid state.[7] It makes various predictions for this region which can be experimentally tested. This includes specific critical temperature dependences, involving a critical temperature $T_c$, expected for various dynamic properties. However, close to $T_c$ and below, additional effects are commonly believed to come into play, like the mentioned hopping processes treated by extended MCT, the influence of energy landscapes, or the cooperativity/heterogeneity of molecular motions.[3,4,5,8,9,10]

There are many liquids (sometimes termed "good glass formers") that can be supercooled and transformed into a glass by applying easily achievable cooling rates. Examples are silicates as used, e.g., for windows or bottles, many polymers, some ionic liquids and salt melts, and various molecular glass formers, e.g., glycerol. MCT was mainly applied to such good glass formers, and, mostly, their properties were found to be in reasonable accord with the theoretical predictions (see, e.g., Refs. 11,12,13,14,15,16,17). However, in principle, any liquid (except for special cases like helium) can be supercooled and transformed into a glass, although in many cases this is technically difficult due to the exceedingly high cooling rates required to avoid crystallization. Maybe the most prominent example is water, which can be considered as the most important liquid on earth. It can be only slightly supercooled by conventional methods (down to about -38 °C or 235 K), can be prepared in a stable amorphous and solid state only below about 150 K (usually in the form of thin films or tiny droplets), and inevitably crystallizes when cooling or heating it into the temperature region between 150 and 235 K, the so-called no-man's land (NML) of water.[18,19,20,21]

As MCT should be applicable to all liquids,[7] naturally the question arises whether liquid water, being of outstanding importance in nature and technology, also follows its predictions. As mentioned, the latter concern various dynamical properties such as the temperature-dependences of the viscosity $\eta$ or the relaxation time $\tau_\alpha$, characterizing molecular dynamics. MCT also predicts various peculiarities of the relaxation dynamics as revealed, e.g., by susceptibility spectra, which are expected to comprise several contributions with different time scales. Unfortunately, for water these predictions are difficult to test within the stable amorphous region below the NML (i.e., at $T < 150$ K), as here it is usually available in the form of thin films or otherwise restricted geometries only. Especially, dielectric measurements in the GHz–THz frequency range, where MCT expects the appearance of peculiar fast dynamics, are difficult to perform for such samples. Moreover, often a liquid-liquid transition or crossover is assumed for water within the NML, sometimes termed "fragile-to-strong transition" (FST).[19,20,21,22,23] This implies that water's properties at temperatures below the NML may not be representative of its more common form as existing at room temperature and ambient pressure. Thus, studies of the applicability of MCT to liquid water are restricted to high temperatures and to temperatures that advance as far as possible into the NML from above.

Until now, such investigations were mainly performed using scattering, rheological, and nuclear magnetic resonance (NMR) measurements.[24,25,26,27,28,29,30,31] Moreover, several works also tackled this question via molecular-dynamics (MD) simulations.[32,33,34,35,36,37,38,39] Most of these studies analyzed the temperature dependence of the $\alpha$-relaxation time $\tau_\alpha(T)$, for which MCT predicts a critical law. It is supposed to characterize the dominant molecular motions that govern the glassy freezing. Although dielectric spectroscopy is a prominent method for the determination of $\tau_\alpha$ of molecular liquids and was often used to check the applicability of MCT in other liquids,[14,15,17,40,41,42] to our knowledge it has never been employed to investigate water dynamics in terms of MCT. Moreover, the validity of MCT's detailed predictions concerning the properties of the so-called fast $\beta$ relaxation and its close entanglement to the $\alpha$ relaxation have only rarely been checked until now, mostly by applying light and neutron scattering to bulk[24] and confined water.[25,26]

In Ref. 43, we have reported nearly continuous broadband dielectric spectra of liquid water of unprecedented precision, measured by a combination of experimental methods. In the present work, we check what can be learnt from these data about the possible applicability of MCT to water. As crystallization in the NML can be prevented by admixing salts to water,[44,45,46,47] we also analyze recent broadband dielectric data on $H_2O$:LiCl solutions from our group[21,43] in terms of MCT and discuss their relevance for pure water. Moreover, we treat the question whether information on the controversial FST of water can be obtained from a MCT analysis of the available dielectric $\tau_\alpha(T)$ data on liquid and slightly supercooled pure water. Interestingly, evidence for such an FST was recently reported, based on the application of MCT to relaxation-time data from ultrafast x-ray measurements, monitoring structural relaxation in micrometer-sized water droplets.[48] We also compare these and the present dielectric $\tau_\alpha$ data to those from further experimental methods. Finally, we briefly discuss the problem of contradicting interpretations of water's main relaxation process as detected by dielectric spectroscopy[49,50,51,52,53,54] and their implications for the present MCT analysis. For simplicity reasons, in the following this process will be denoted $\alpha$ process, although this assignment is disputed.

## II. MCT PREDICTIONS

When applying MCT to dielectric data, usually spectra of the loss, $\varepsilon''(\nu)$, are analyzed, in analogy to the often-treated susceptibilities $\chi''$ detected by scattering experiments.[16,17,42] In most molecular liquids, the observed main loss peak is ascribed to the $\alpha$ relaxation. Then the found many-decades shift of its peak frequency $\nu_\alpha$ to lower frequencies with decreasing temperature mirrors the slowing down of molecular motion when the glass transition temperature is approached [see schematic plot of typical loss spectra at two different temperatures in Figs. 1(a) and (b)]. Aside of predictions concerning the temperature dependence of $\nu_\alpha(T)$ [or of $\tau_\alpha(T) \approx 1/(2\pi\nu_\alpha)$], MCT also makes clear statements about the emergence of the so-called fast $\beta$ relaxation, ascribed to particle rattling within the transient cage formed by surrounding molecules.[2] It should lead to a shallow loss minimum at higher frequencies than the $\alpha$ peak (mostly at 10–100 GHz[12,13,14,15,16,17,41,42]), which is expected to arise in excess of contributions from even faster processes within the THz region, see Fig. 1(b). In the dielectric-spectroscopy literature, the most prominent among these faster processes is reported in the region around 1 THz and often termed boson peak (BP),[17,40,55,56,57] analogous to corresponding spectral features arising in light- and neutron-scattering spectra.[13,58] Its microscopic origin is still controversially discussed. In Ref. 43, we proposed it to also contribute to broadband dielectric-loss spectra of water, albeit strongly superimposed by the dominating $\alpha$-relaxation peak. At even higher frequencies, resonance modes, e.g., due to intramolecular excitations show up [marked by "res" in Figs. 1(a) and (b)].

As will be detailed below, MCT makes distinct predictions concerning the spectral shape and temperature dependence of the loss minimum, whose properties should be closely entangled with those of the $\alpha$ relaxation, occurring at much lower frequencies.[2] Unfortunately, at high temperatures, in the low-viscosity liquid where the $\alpha$-relaxation rate reaches GHz frequencies and beyond, the minimum becomes obscured by the overlap of the $\alpha$ and boson peaks [Fig. 1(a)]. This was found in various dielectric investigations of molecular glass formers[17,40,41,42] and also in susceptibility spectra derived from scattering experiments.[12,16,59,60] Therefore, the mentioned restriction of water measurements to rather high temperatures, avoiding the NML, represents a major problem, making it virtually impossible to analyze the susceptibility minimum of pure water in terms of MCT.

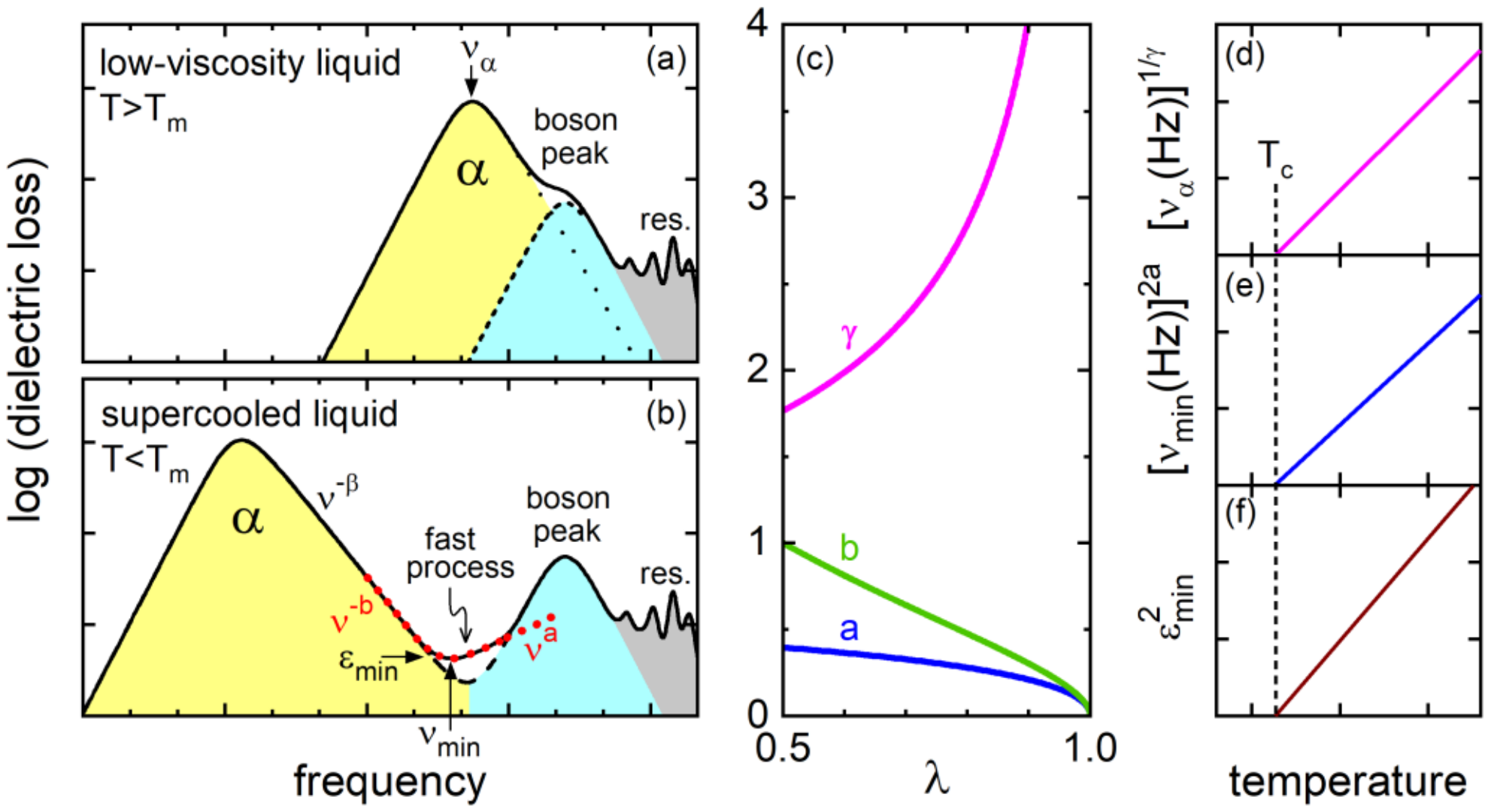

**FIG. 1.** (a) Schematic dielectric-loss spectrum of a dipolar liquid in its low-viscosity state at high temperatures above the melting point $T_m$. Different dynamic processes are indicated by shaded areas. (b) The same for the supercooled state at a temperature significantly below $T_m$. The loss minimum predicted by MCT (dotted red line) and its relevant parameters are indicated, assuming the simplest case, $b = \beta$. The dashed black line shows the hypothetical spectrum in the minimum region without the fast $\beta$ process.[17,42] For this plot, the temperature is assumed to be sufficiently high to exclude the occurrence of excess wings or Johari-Goldstein relaxations which often appear at lower temperatures. (c) Dependence of the critical MCT exponents on the MCT system parameter $\lambda$ as derived from Eqs. (3) and (4). (d)–(f) Critical laws predicted by MCT for the temperature-dependent $\alpha$-peak frequency [Eq. (1)], minimum frequency [Eq. (5)], and minimum loss amplitude [Eq. (6)], respectively, using representations that lead to linear behavior. The vertical dashed line indicates $T_c$.

In the following, we briefly summarize the predictions of idealized MCT that are relevant for the present work:[2] $\nu_\alpha(T)$ [and, thus, $\tau_\alpha(T)$] should follow a critical law with exponent $\gamma$.

$$\nu_\alpha \propto (T - T_c)^{\gamma} \text{ or } \tau_\alpha \propto (T - T_c)^{-\gamma} \qquad (1)$$

Above $T_c$, time-temperature superposition (TTS) should hold for the $\alpha$ relaxation, implying a temperature-invariant peak shape. Below $T_c$, a square-root behavior of the temperature-dependent relaxation strength is expected, which, however, cannot be tested in water because the NML prevents measurements at low temperatures.

The predicted high-frequency loss minimum can be described by a sum of two power laws with exponents $-b$ (left flank) and $a$ (right flank):

$$\varepsilon'' = \frac{\varepsilon_{\min}}{a+b}\left[a\left(\frac{\nu}{\nu_{\min}}\right)^{-b} + b\left(\frac{\nu}{\nu_{\min}}\right)^{a}\right] \qquad (2)$$

Here $\varepsilon_{\min}$ and $\nu_{\min}$ are the minimum amplitude and frequency, respectively as indicated in Fig. 1(b). The two exponents should be temperature independent and directly linked to each other via the system-dependent parameter $\lambda$:

$$\lambda = \frac{\Gamma^2(1-a)}{\Gamma(1-2a)} = \frac{\Gamma^2(1+b)}{\Gamma(1+2b)} \qquad (3)$$

where $\Gamma$ is the gamma function. In the simplest case, the low-frequency flank of the minimum is identical with the high-frequency flank of the $\alpha$ peak which usually follows a power law $\nu^{-\beta}$ for frequencies sufficiently beyond the peak frequency. This implies $\beta = b$ as schematically indicated in Fig. 1(b). This scenario only applies when so-called excess wings or Johari-Goldstein secondary relaxations can be neglected, which often is the case at high temperatures. In contrast, upon cooling, in most liquids these spectral features start to superimpose the right $\alpha$-peak flank.[17,42,61,62,63] However, they do not seem relevant for water within the available temperature range.[43]

Remarkably, according to MCT the exponents $a$ and $b$, defining the minimum's spectral form, should also determine the temperature dependence of the $\alpha$ relaxation, which is located at much lower frequencies. Namely, the exponent $\gamma$ in Eq. (1) is given by:

$$\gamma = 1/(2a) + 1/(2b). \qquad (4)$$

As $\beta = b$ and as $a$ can be calculated from $b$ via Eq. (3), the right flank of the $\alpha$ peak ($\varepsilon'' \propto \nu^{-\beta}$) and the temperature dependence of its peak frequency $\nu_\alpha$ [Eq. (1)] should be closely linked. Fig. 1(c) depicts the dependence of $a$, $b$, and $\gamma$ on $\lambda$ as resulting from Eqs. (3) and (4). As MCT predicts $\lambda > 0.5$ and as $b = \beta$ cannot be larger than one, $a$ should never exceed 0.395.[2,16,64] Consequently, there is a theoretical lower limit for $\gamma$, which cannot become smaller than 1.76.

For the temperature dependence of the minimum, MCT also predicts critical laws, namely:[2]

$$\nu_{\min} \propto (T - T_c)^{1/2a} \qquad (5)$$

$$\varepsilon_{\min} \propto (T - T_c)^{1/2} \qquad (6)$$

Notably, according to Eq (5), the temperature dependence of $\nu_{\min}$ is governed by the exponent $a$ of the right wing of the minimum. Moreover, $a$ itself is related to the minimum's left-wing slope $-b$ and, thus, to $-\beta$ determining the right $\alpha$-peak flank. The three critical laws of Eqs. (1), (5), and (6) are schematically indicated in Figs. 1(d)–(f), using representations that should linearize the experimental data. Then $T_c$ (dashed line) can be read off by linear extrapolation to an ordinate value of zero.

## III. RESULTS AND DISCUSSION

### A. Pure water

Figure 2(a) presents typical spectra for pure water,[43] measured by combining data from "classical" dielectric spectroscopy, coaxial reflectometry,[65,66] a quasi-optical Mach-Zehnder setup,[67] a terahertz time-domain spectrometer, and a Fourier-transform infrared spectrometer. In contrast to previously published data, they cover different temperatures, are nearly continuous, and were collected by a single workgroup only, using identical sample preparation and identical temperatures at all frequency ranges. Moreover, the often-neglected real part of the permittivity $\varepsilon'$ was also determined,[43] which significantly enhances the precision of the analysis.

The experimental data of Fig. 2(a) approximately correspond to the situation depicted in the schematic spectrum in Fig. 1(a). Especially, the fast $\beta$ process is hidden by the overlapping contributions from the $\alpha$ relaxation and boson peak because in water the accessible temperatures above the NML are too high, as discussed in the preceding section. Nevertheless, the main loss peak, at room temperature located at ~20 GHz and termed $\alpha$-relaxation peak here, is fully resolved. This allows for unequivocal conclusions on its peak frequency *and* spectral shape. As discussed in detail in Ref. 43, this shape reveals small but significant deviations from the Debye function. The latter is predicted by the Debye theory of dielectric relaxation[68] and, in the log-log representation of Fig. 2(a), would lead to a slope of $-\beta = -1$ at the high-frequency peak flanks. Instead, the empirical Cole-Davidson (CD) function[69] [dashed lines, shown for 275 and 350 K in Fig. 2(a)] provides a better description of the measured spectra. It allows for more shallow high-frequency slopes and was found to be applicable to the $\alpha$ relaxation of many dipolar liquids.[17,40,69,70] For water, it captures the right peak flanks down to much lower $\varepsilon''$ values than fits with the Debye function which are indicated by the solid lines in Fig. 2(a).[43] This is even more clearly confirmed by the simultaneous analysis of the corresponding $\varepsilon'$ spectra[43] as demonstrated for 350 K in the inset of Fig. 2(a). Such peak broadening, compared to the Debye case, is usually explained by a disorder-induced distribution of relaxation times.[71,72]

The value $\beta = 0.9$, obtained for water at 275 K from the CD fits, is not too far from the Debye value $\beta = 1$. This may be a reason for the widespread claim of Debye shape for the main dielectric process of pure water.[50,52,73,74,75,76,77,78,79,80] It seems to be founded on pioneering dielectric investigations of water where low loss values at high frequencies were difficult to detect,[81,82,83] albeit in fact some early hints at non-Debye behavior were published.[84,85] While a free fit of the 350 K curve leads to $\beta = 0.95$,[43] it also can be reasonably well fitted with $\beta$ fixed to 0.9 as obtained for 275 K [right dashed line in Fig. 2(a)]. This finding is in accord with TTS predicted by MCT for high temperatures, $T > T_c$. We consider $\beta = 0.9$ as more reliable because, at 275 K, the $\alpha$ peak is located at lowest frequencies, and, thus, its high-frequency flank is least influenced by high-frequency processes like the suggested BP[43] or resonances, both not covered by MCT.

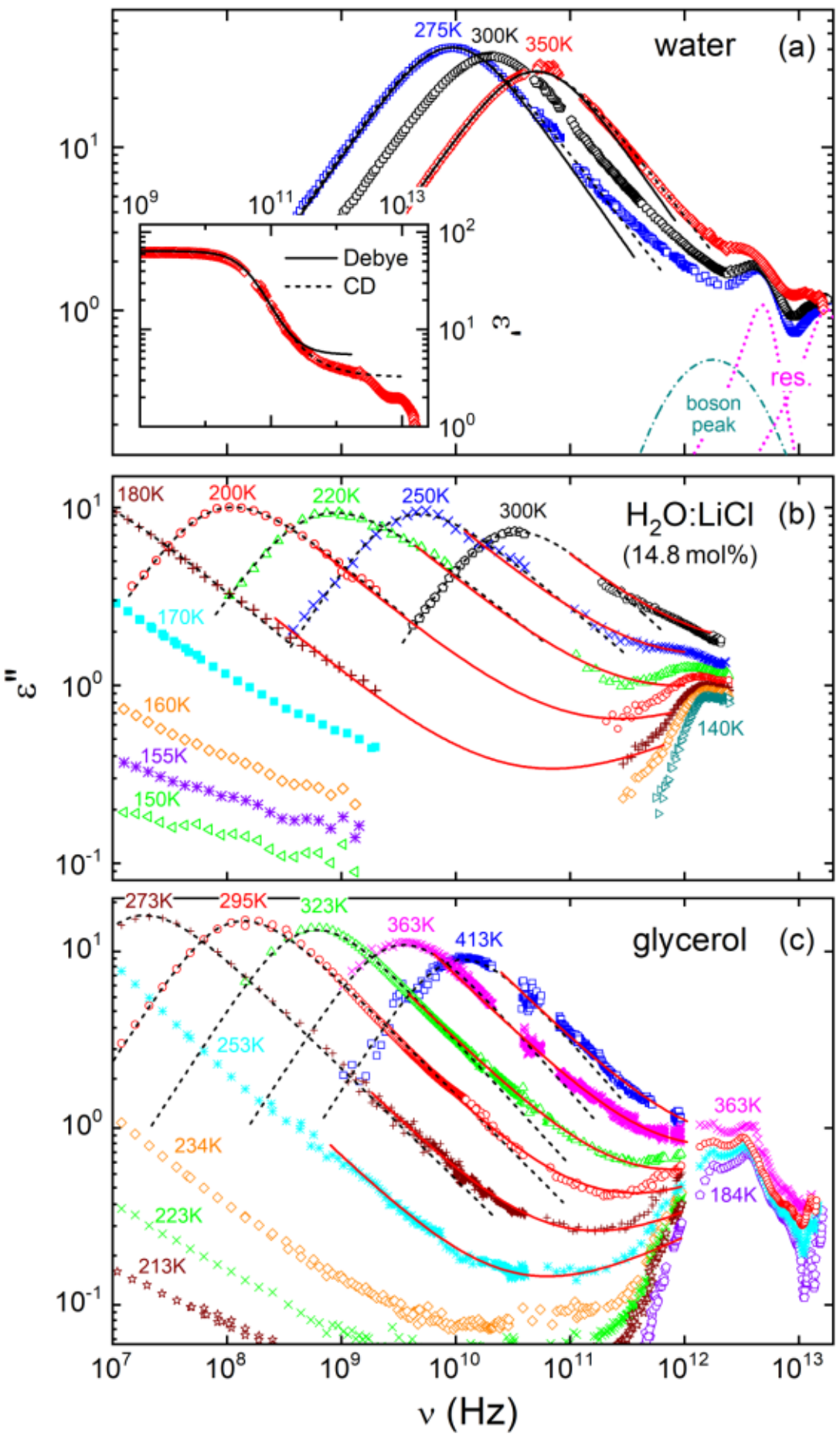

**FIG. 2.** Dielectric loss spectra of three dipolar liquids for frequencies from 10 MHz well into the THz range. (a) Symbols: pure water at three typical temperatures.[43] Dashed and solid lines: fits of the $\alpha$ relaxation with the CD or Debye function, respectively. Dash-dotted and dotted lines: high-frequency contributions to the 275 K spectrum as suggested in Ref. 43. The inset in (a) presents the $\varepsilon'$ spectrum of water at 350 K[43] with corresponding fits of the $\alpha$ relaxation. (b) Symbols: spectra of a 14.8 mol% aqueous LiCl solution, corrected for the contributions from charge transport.[43,91] Dashed lines: fits of the $\alpha$ peaks with the Havriliak-Negami function (see Ref. 43 for details). (c) Symbols: spectra of glycerol.[89,90] Dashed lines: CD fits of the $\alpha$ peaks. The solid red lines in (b) and (c) were calculated with the minimum formula of MCT, Eq. (2), as explained in the text.

When fitting the $\alpha$ peak of water with the Debye function, implying $\beta = 1$ [solid lines in Fig. 2(a)], deviations from the experimental data already arise at frequencies not far beyond $\nu_\alpha$, although the right peak flank follows a power law without any discontinuity up to much higher frequencies. This leads to the impression of a rather strong excess contribution to $\varepsilon''$, starting to show up less than one frequency-decade beyond $\nu_\alpha$. In later literature (e.g., Refs. 50, 73, 74, and 79), covering higher frequencies than the mentioned pioneering works, it is often accounted for by including one or more additional Debye peaks in the overall fit functions (see Ref. 43 for a more detailed discussion). However, even when postulating that the $\alpha$ peak of water must have Debye shape for some reason, the assumption of a second Debye peak at its right flank neglects the fact that secondary processes in dipolar liquids are always significantly broadened.[86,87] As will be discussed in more detail in section III.D, also alternative scenarios are considered where water's $\alpha$ relaxation is not reflected by its main loss peak but, instead, only leads to the mentioned excess contributions at the right flank of the latter.[51,52,53] However, then again the description of these contributions by a second Debye peak may be questioned because $\alpha$-relaxation peaks commonly are significantly broader.

Interestingly, when fitting the main peak of water with the CD function [dashed lines in Fig. 2(a)], at high frequencies and for $\varepsilon''$ far below its peak value, still deviations of fit and data show up, albeit of significantly weaker amplitude than for fits with the Debye function. In Ref. 43, the BP was suggested to explain these remaining deviations, together with the well-known vibration-related resonance processes.[75,77,78,79] An alternative explanation in terms of the coupling model was proposed in Ref. 88. As an example for the BP scenario, the dash dotted line in Fig. 2(a) indicates a rough estimate of the BP of water at 275 K.[43] We want to point out that the overall findings for the dielectric-loss spectra in the accessible temperature range of liquid water resemble those in other dipolar liquids at high temperatures. An example is glycerol at 363 or 413 K as shown in Fig. 2(c).[89,90] As schematically indicated in Fig. 1(a) and (b), there the BP only becomes visible when, upon cooling, the $\alpha$ peak shifts to lower frequencies, which for glycerol is the case at $T \leq 323$ K [Fig. 2(c)]. Figure 2(b) reveals that the situation in a 14.8 mol% aqueous LiCl solution[43,91] is similar. In contrast to pure water, here supercooling is possible, uncovering the smooth development of the mentioned excess contribution at the right $\alpha$-peak flank (seen at 300 K) into a well-resolved BP at about 1–2 THz (clearly revealed below ~200 K). Moreover, despite some deficits concerning the continuity of the spectra, upon cooling here the emergence of a $\varepsilon''$ minimum becomes obvious, just as for glycerol [Fig. 2(c)] and other dipolar liquids.[14,17,41,42,87,90] We will treat these data in more detail in section III.B. Finally, it should be noted that the spectral feature termed "excess wing",[17,61,89] known to arise in many supercooled liquids, cannot explain the high-frequency deviations of the water spectra from the CD fits, treated in this paragraph. It only shows up at low temperatures and is not observed when the $\alpha$-peak frequency is in the 10 GHz range as for water above the NML.[17,42,43]

As discussed in Ref. 21, from fits of the nearly continuous $\varepsilon'$ and $\varepsilon''$ spectra of Ref. 43, covering the $\alpha$ relaxation both below and above the peak frequency at various temperatures, we determined the average main relaxation times of water with high precision. While, with some effort, bulk water in principle can be supercooled down to about 235 K, such supercooling was not possible with the experimental setups employed in Ref. 43. We thus resort to data from literature to expand the $\tau_\alpha(T)$ data analyzed in the present work to temperatures below 273 K. Among the various reported dielectric $\tau_\alpha$ data sets of water (see Fig. 7 in Ref. 21 for examples),[74,92,93,94] we found that those from Bertolini *et al.*[92] provide the most appropriate extension of our own results towards low temperatures. They agree well with those reported in Refs. 74 and 93,[21] but extend down to smaller temperatures. The combined $\tau_\alpha$ data are shown in Fig. 3 by the crosses[21] and plusses[92] using an Arrhenius representation. The stars in Fig. 3(a) represent experimental values of $\tau_\alpha$ at temperatures below the NML (indicated by the shaded area), reported for deeply supercooled "low-density liquid" water, by Ammann-Winkel *et al.*[95] It is clear that a continuous connection of the data below and above the NML is impossible, which is discussed in detail in Ref. 21. Especially in the magnified view of Fig. 3(b), the well-known non-Arrhenius behavior of $\tau_\alpha(T)$ of water above the NML is clearly revealed by the non-linear curve shape. The inset of Fig. 3 provides a comparison with the $\eta(T)$ data sets of water from Refs. 96, 97, and 98 (three closely overlapping solid lines), together covering a broad temperature range. These data were vertically scaled to match $\tau_\alpha(T)$.[99] The reasonable agreement indicates that the dynamics causing the dominant relaxation process in dielectric water spectra, termed $\alpha$ relaxation in the present work, is closely related to the structural molecule dynamics governing viscosity. We will discuss this issue in more detail in section III.D.

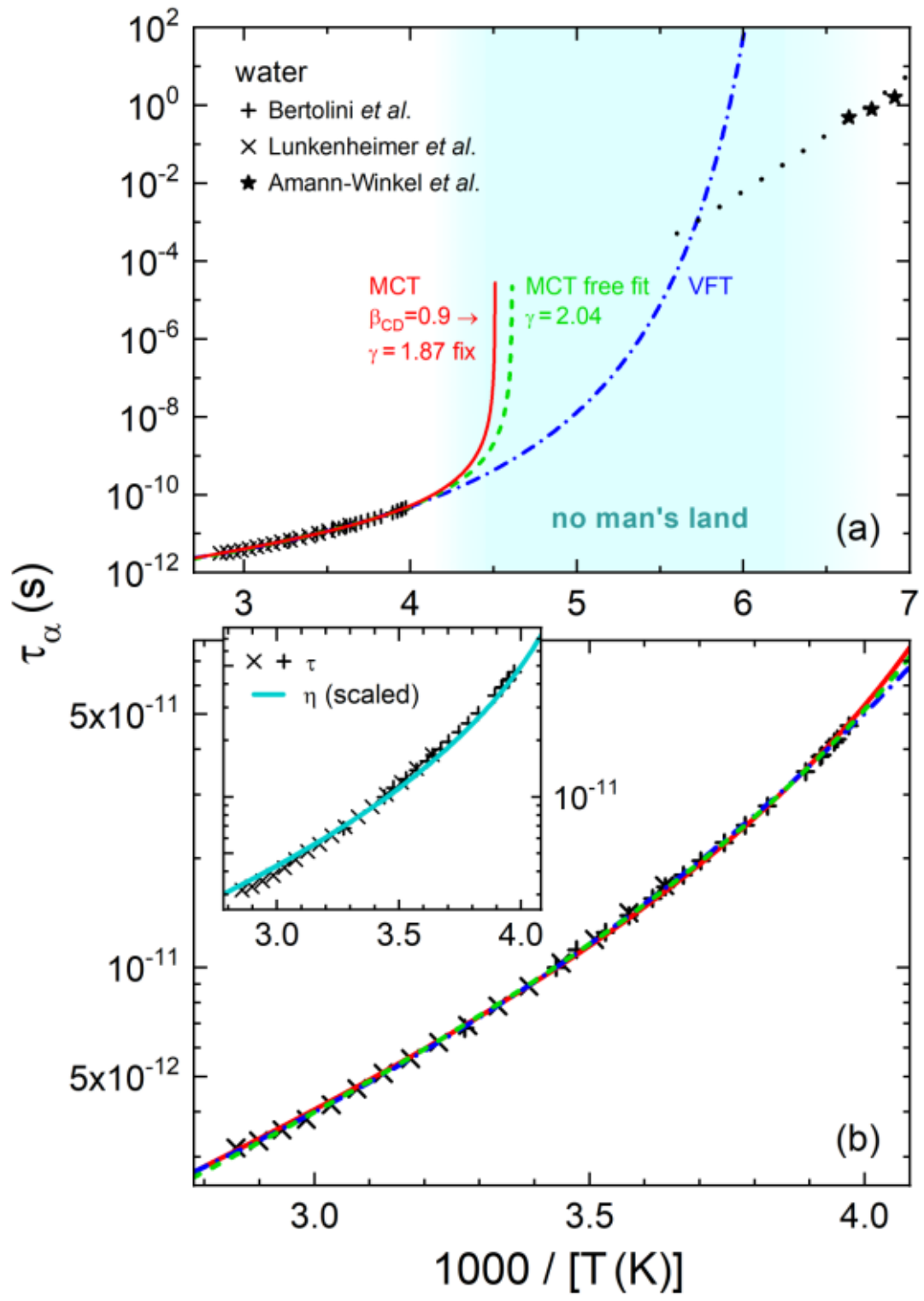

**FIG. 3.** (a) Temperature dependent relaxation time of the main dipolar relaxation process in pure water shown using an Arrhenius representation. The data above the NML (shaded region) are from Refs. 21 (×) and 92 (+). Those below the NML were taken from Ref. 95. The lines (except the dotted one) show fits of the data above the NML. Dash-dotted: VFT [Eq. (7)]; dashed: MCT critical law, Eq. (1), with free parameters; solid: MCT with $\gamma$ fixed to 1.87, corresponding to $b = \beta = 0.9$ [cf. Eqs. (3) and (4)] where $\beta$ was deduced from Fig. 2(a). The dotted line in (a) shows $\tau(T)$ of the strong form of water as suggested in Ref. 21. (b) Magnified view of the high-temperature region in (a). The inset shows $\tau(T)$ above the NML (symbols) together with the viscosity $\eta(T)$ (three lines appearing as a single one due to their close overlap),[96,97,98] which was vertically scaled[99] to match the dielectric data.

The dash-dotted line in Fig. 3(a) shows a fit[21] of $\tau_\alpha(T)$ above the NML with the phenomenological Vogel-Fulcher-Tammann (VFT) formula,[100,101,102,103]

$$\tau_\alpha = \tau_0 \exp\left(\frac{DT_{\mathrm{VF}}}{T - T_{\mathrm{VF}}}\right) \quad (7),$$

often employed to describe non-Arrhenius behavior in glass forming liquids.[3,4,17,42,104] Here $\tau_0$ is a pre-exponential factor, the so-called strength parameter $D$ determines the degree of deviation from Arrhenius temperature dependence,[103] and $T_{\mathrm{VF}}$ is the Vogel-Fulcher temperature, where $\tau_\alpha$ is predicted to diverge. The dotted line at low temperatures in Fig. 3(a) is the proposed behavior of strong water at temperatures below the FST deduced in Ref. 21 from measurements of aqueous solutions. Based on this analysis, in Ref. 21 the FST of pure water was estimated to occur at roughly 175 K.

One should note that in Ref. 48 a much higher FST temperature of 233.2 K was reported, based on fits of relaxation-time data between 228 and 270 K which were deduced from ultrafast structural-relaxation measurements on tiny water droplets. For these fits, the critical law of MCT, Eq. (1), was employed, which was found to be valid at $T > 233.2$ K, below which a crossover of $\tau_\alpha(T)$ to Arrhenius behavior was stated. In MCT analyses of supercooled liquids, indeed $\tau_\alpha(T)$ is often fitted by Eq. (1) at temperatures above $T_c$, while for lower temperatures, where idealized MCT is no longer valid, other functions are used.[40,41,105,106,107] Such an approach is also in accord with the random first-order transition (RFOT) theory,[108,109] which essentially implies the approximate validity of the critical MCT law above and of the VFT function below $T_c$.[110]

To check for such a scenario in the present dielectric data, the solid and dashed lines in Fig. 3(a) show alternative fits of $\tau_\alpha(T)$ of water above the NML, employing the MCT expression, Eq. (1), instead of the VFT function. As discussed in section II, according to MCT $\gamma$ should be directly related to the right flanks of the $\alpha$-relaxation peaks via Eqs. (3) and (4). Using $b = \beta = 0.9$, derived above from Fig. 2(a), a critical exponent $\gamma = 1.87$ is deduced, corresponding to a system parameter $\lambda = 0.55$ [cf. Fig. 1(c)]. The fit of $\tau_\alpha(T)$ with $\gamma$ fixed to this value, shown by the solid lines in Figs. 3(a) and (b), is of good quality and leads to $T_c = 222$ K. This agrees reasonably with reported critical temperatures of water between 220 and 227 K determined using different experimental methods.[24,27,28,29,30,31] As an additional check, the dashed line in Fig. 3 represents a fit with Eq. (1) where all parameters are free, leading to $T_c' = 217$ K and $\gamma' = 2.04$, the latter corresponding to $b' = 0.78$. In literature,[24,27,28,29,30,31] $\gamma$ values of water from free fits varying between 1.6 and 2.3 were reported, except for Ref. 48, where $\gamma = 0.82$ was found (we will discuss this deviation in section III.C). The closeness of $T_c'$ from the free fit to $T_c = 222$ K, as obtained from the fit with $\gamma = 1.87$ derived from $b = \beta = 0.9$, provides confidence in the deduced absolute value of $T_c = 222$ K.

As most clearly revealed in the magnified view of Fig. 3(b), the MCT fits are of comparable quality as those with the VFT function. Thus, we conclude that the dielectric $\tau_\alpha(T)$ data above the NML of pure water are well consistent with MCT. Remarkably, this statement also holds when considering the predicted interconnection of the temperature dependence of $\tau_\alpha$ and the loss-peak shape, which fixes the value of $\gamma$ in the fits with Eq. (1). It should be noted that these dielectric data on bulk water reveal no indications of a crossover to Arrhenius behavior at low temperatures. However, one should be aware that they only cover temperatures down to about 252 K while those in Ref. 48 on tiny water droplets extend down to 228 K.

## B. LiCl solutions

Spectra on aqueous solutions are often employed to learn more about the properties of supercooled water because their reduced crystallization tendency causes a smaller or even completely absent NML.[21,44,45,46,47] The broadest dielectric spectra are available for a 14.8 mol% LiCl solution[21,43,91] for which typical loss spectra above 10 MHz are shown in Fig. 2(b). In contrast to the mentioned loss minima, hampered by frequency gaps lacking experimental data points, the peak frequencies and right flanks of the $\alpha$ relaxation are mostly well defined for this solution. Notably, they can be observed down to much lower temperatures than for pure water, as no NML exists for this solution. Lower temperatures are also accessible for water with 7.3mol% LiCl, investigated by us in Refs. 21 and 43, where the NML at least is less extended than in water. The temperature-dependent relaxation times deduced for the two concentrations are shown in Fig. 4(a) (Arrhenius representation), together with those for pure water, included for comparison. Figure 4(b) presents a magnified view of the high-temperature region. The solid lines show MCT fits of $\tau_\alpha(T)$ at high temperatures with Eq. (1), where $\gamma$ is fixed to values derived via Eqs. (3) and (4) from $b = \beta$. The latter is again deduced from the right $\alpha$-peak flanks assuming temperature-independent $\beta$. The found $b = \beta = 0.53$ for 14.8 mol% and 0.66 for 7.3 mol% LiCl lead to $\gamma = 2.63$ and 2.27, respectively (Table I). Again, the fits with Eq. (1) and fixed $\gamma$ are of good quality.

**TABLE I.** MCT critical temperature $T_c$, exponents $\gamma$, $\alpha$, and $\beta$, and system parameter $\lambda$, as deduced for pure water and aqueous LiCl solutions. Note that $\gamma$, $a$, $b$, and $\lambda$ are connected via Eqs. (3) and (4) as depicted in Fig. 1(c).

| $x$ (mol%) | $T_c$(K) | $\gamma$ | $a$ | $b$ | $\lambda$ |
|---|---|---|---|---|---|
| 0 | 222 | 1.87 | 0.38 | 0.9 | 0.55 |
| 7.3 | 215 | 2.27 | 0.33 | 0.66 | 0.69 |
| 14.8 | 194 | 2.63 | 0.30 | 0.53 | 0.77 |

As mentioned, $\tau_\alpha(T)$ of the solutions is available down to lower temperatures than for pure water. It is known from other glass forming liquids[40,41,89,106,107,111] that Eq. (1) is unable to fit $\tau_\alpha(T)$ within a large temperature range, extending deep into the supercooled regime. As discussed above, this is expected because idealized MCT is only valid above $T_c$. Therefore, MCT fits as shown in Fig. 4 always require a compromise between sufficiently high fit quality and an as large as possible temperature range. Extrapolating these fit curves in Fig. 4(a) (solid lines) to low temperatures, reveals their divergence at $T_c$ [cf. Eq. (1)]. This divergence, of course, is incompatible with the experimental data as the fits are only valid and were only performed at high temperatures.

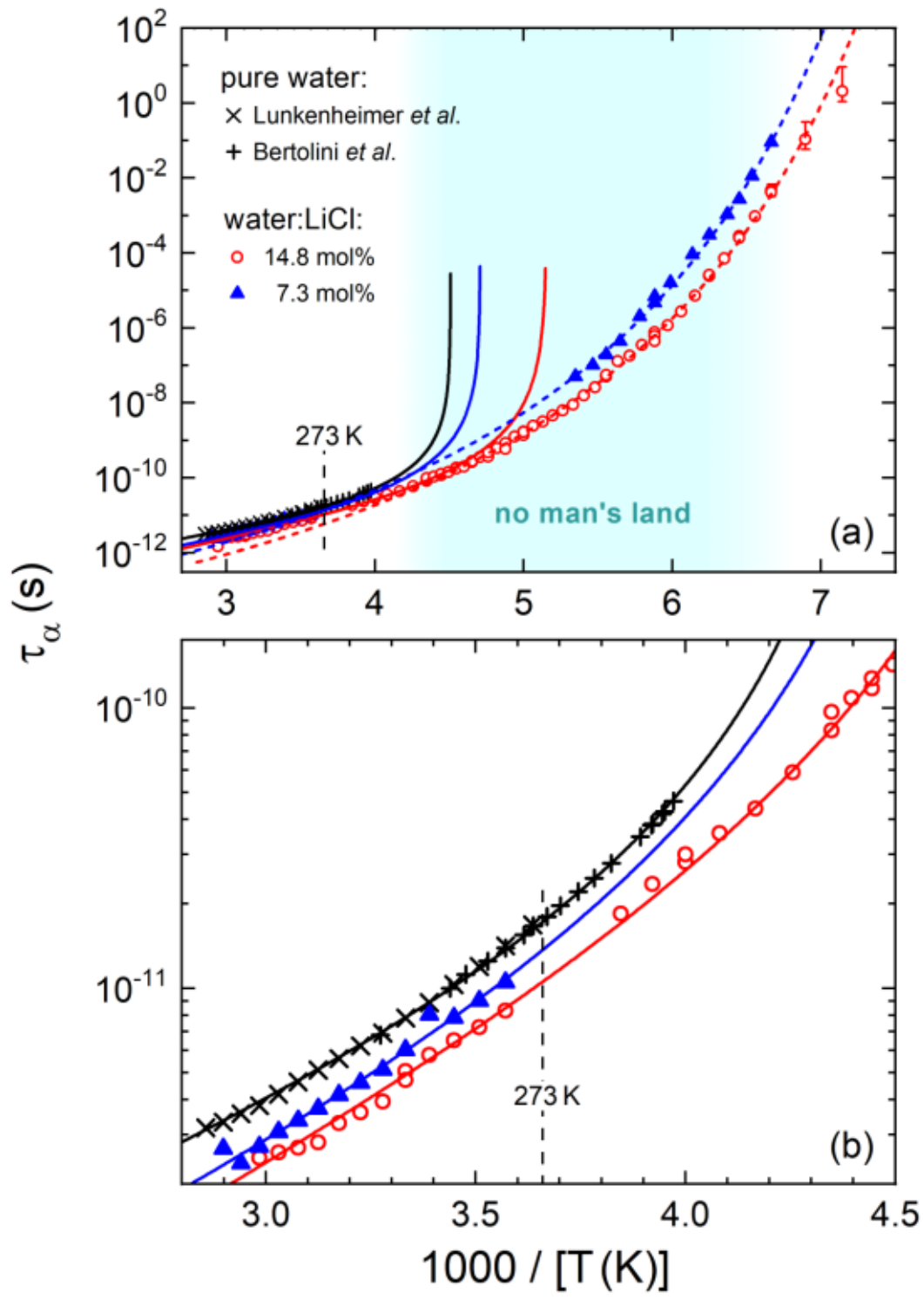

**FIG. 4.** (a) Relaxation times of the main dipolar relaxation processes in pure water[21,92] and two aqueous LiCl solutions[21] as specified in the legend, plotted in Arrhenius representation. The shaded region indicates the NML of pure water. The solid lines show fits of $\tau_\alpha(T)$ at high temperatures with the MCT critical law, Eq. (1) (see Table I for the parameters), where the exponent $\gamma$ was fixed in accord with the values for $b = \beta$ found in Fig. 2 [cf. Eqs. (3) and (4)]. The dashed lines are fits with the VFT function, Eq. (7), performed at low temperatures, excluding the experimental data fitted by Eq. (1). (b) Magnified view of the high-temperature region in (a).

An unequivocal check of the applicability of the critical MCT law, Eq. (1), not hampered by a proper choice of the fitted temperature range, is a plot of the quantity $\nu_\alpha^{1/\gamma}$ vs $T$. This should linearize the experimental data in the range where Eq. (1) is valid, as schematically indicated in Fig. 1(d). Figures 5(a)–(c) show such plots for the 14.8 and 7.3 mol% solutions, and for pure water, respectively. Indeed, the data above $T_c$ are almost perfectly linearized in this way. One should note that, in contrast to $T_c$, here $\gamma$ again is not a free parameter but was directly derived from the right flanks of the experimental loss peaks. The lines in Figs. 5(a)–(c) correspond to the same fit curves by Eq. (1) as already shown in Fig. 4 (solid lines). The resulting $T_c$ values (Table I) can be read off from their intersection with the abscissa. The deviations of the experimental low-temperature data from linear behavior in this representation, as observed for the solutions, is due to the mentioned fact that the MCT critical laws are only valid above, but not too close to $T_c$. This is in accord with the low-temperature deviations of fits and experimental data in Fig. 4(a), discussed above. Figures 5(d) and (e) show the development of $T_c$ and $\gamma$ with the LiCl concentration $x$ as resulting from the above analyses. The found smooth, nearly linear behavior looks reasonable and supports the validity of the determined $T_c = 222$ K of pure water.

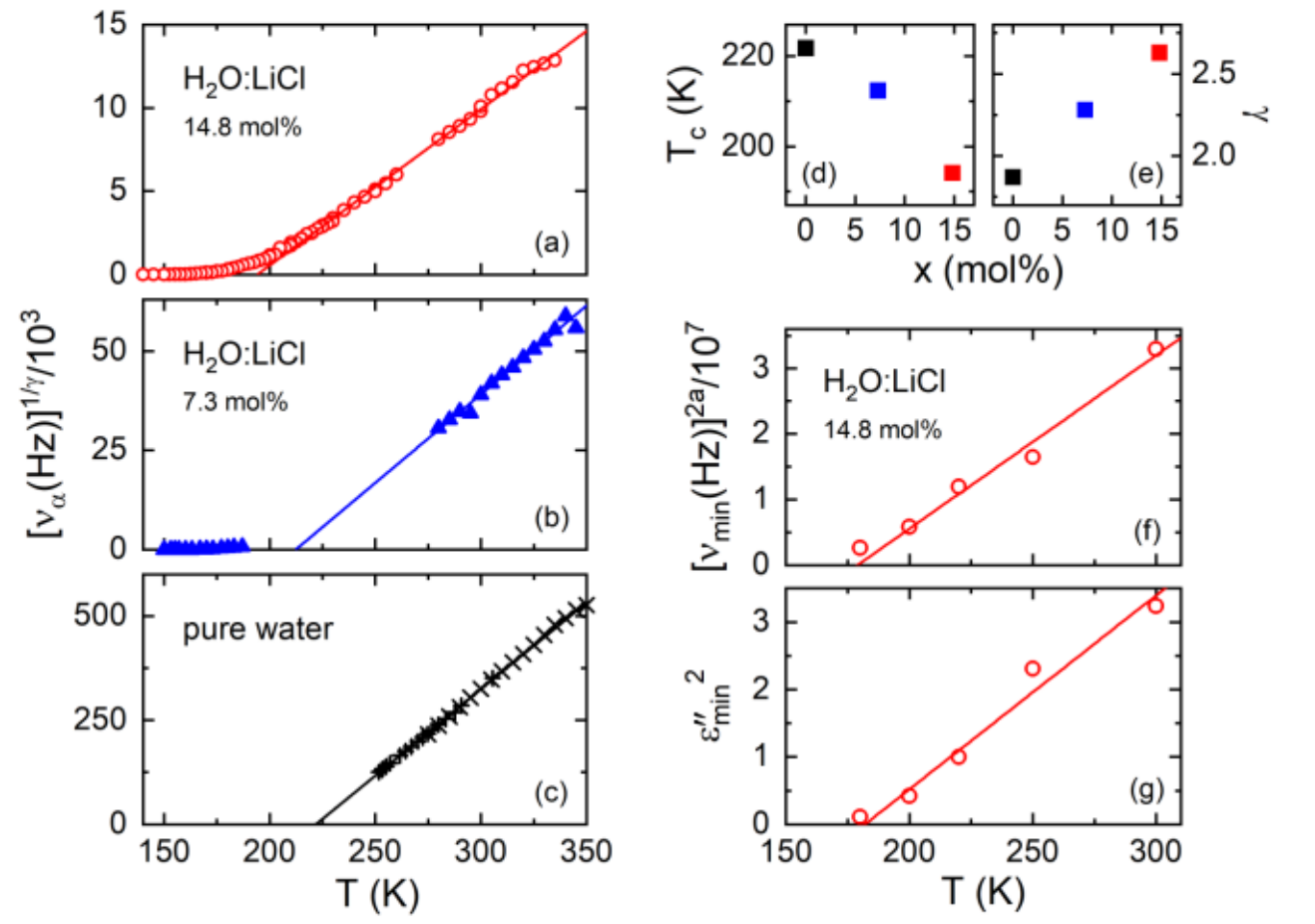

**FIG. 5.** (a)-(c): Temperature dependence of the $\alpha$-relaxation rates $\nu_\alpha$ of the two aqueous LiCl solutions and pure water. The chosen plot of $\nu_\alpha^{1/\gamma}$ vs $T$ (with $\gamma$ fixed assuming $b = \beta$, the latter deduced from Fig. 2) should result in linear behavior according to the MCT critical law, Eq. (1). The lines are the same as the solid lines shown in the Arrhenius representation of Fig. 4. (d) and (e): Exponent $\gamma$ (fixed) and critical temperature $T_c$ (free fit parameter), respectively, as resulting for the three investigated liquids with salt concentration $x$. (f) and (g): Minimum frequency $\nu_{min}$ and amplitude $\varepsilon_{min}$, respectively, as estimated for the 14.8 mol% aqueous LiCl solution in Fig. 2(b). The chosen representations should linearize the critical MCT laws, Eqs. (5), and (6). The solid lines in (f) and (g) are linear fits, ignoring the respective data point at the lowest temperature. For parameters, see Table I.

The dashed lines in Fig. 4(a) are fits with the VFT function, Eq. (7), applied to the solution data points at low temperatures, where Eq. (1) of MCT is not applicable. For the nearly continuous $\tau_\alpha(T)$ data of the 14.8 mol% sample, the combination of MCT and VFT function (which is a reasonable approximation within the RFOT model[110]) provides a better fit than the VFT description of the whole temperature range as shown in Fig. 3 of Ref. 21. Of course, this could be expected when considering the larger number of parameters for the combined fits.

As mentioned in the preceding section, the loss spectra of the 14.8 mol% solution in Fig. 2(b) point to the emergence of a minimum upon cooling. The solid lines represent the best description of the data in this region by the MCT minimum formula, Eq. (2), with temperature-independent and interrelated right and left slopes $a$ and $-b$, as predicted by the theory. Due to the large frequency gaps lacking experimental data points, these lines represent a rough estimate only. Nevertheless, at least the slope and amplitude of the low-frequency wings of the minima are well fixed by the right flanks of the $\alpha$ peaks. As mentioned above, a temperature-independent exponent $b = \beta = 0.53$, defining the low-frequency slopes $-b$ of the minima, leads to a good match of the right flanks of the observed $\alpha$ peaks. This is supported by the dashed lines in Fig. 2(b) which are the fits of the $\alpha$ relaxation with the empirical Havriliak-Negami function[112] as reported in Ref. 43. Their high-frequency flanks indeed are well compatible with $b = 0.53$.

Via Eq. (3), this value implies $a = 0.296$, thereby also fixing the slopes at the right flanks of the calculated MCT minima [solid lines in Fig. 2(b)]. However, obviously $a$ is too small to account for the observed increase of $\varepsilon''(\nu)$ at the left flank of the THz peak, interpreted as BP in Ref. 43. Interestingly, a similar situation is found in various other glass formers in literature.[13,17,87,113,114] A prominent example is glycerol, whose loss spectra at high frequencies are shown in Fig. 2(c).[89,90] Here, at high frequencies the fits with Eq. (2) (solid lines) are limited by the additional, steeper increase towards the BP, not taken into account by the MCT expression.[17] Particularly at the higher temperatures, $T \geq 295$ K, the BP of glycerol completely superimposes the predicted $\nu^a$ power law at the right minimum flank. In glycerol, the latter starts to become discernible only below about 273 K, because the minimum shifts to lower frequencies upon cooling, while the BP position remains nearly unchanged. Unfortunately, no quasi-optical measurements[67] (~50–1200 GHz) were made for the LiCl solution,[43] and the employed open-ended coaxial-line probe technique[66] (~100 MHz – 40 GHz) does not work for small $\varepsilon''$ values. Therefore, in contrast to glycerol [Fig 2(c)], the frequency gap in the solution spectra at low temperatures [Fig 2(b)] is too large to reveal the $\nu^a$ critical law. Nevertheless, a comparison of Figs. 2(b) and (c) points to quite similar general behavior within the minimum region. Notably, with $b$ and $a$ fixed, the only free parameters in Eq. (2) are $\nu_{\min}$ and $\varepsilon_{\min}$. Their adjustment corresponds to bare horizontal and vertical shifts of the minima curves, whose shape, according to MCT, should be identical for all temperatures. As seen from the solid lines in Fig. 2(b), we adapted those shifts to make the right minimum flanks match the low-frequency data from the THz measurements, while keeping the left flanks fixed. It is clear, however, that these lines should only be regarded as an educated guess.

Despite the mentioned deficiencies, it seems interesting to check whether the so-obtained minimum positions and amplitudes of the 14.8 mol% LiCl solution are compatible with MCT. The temperature dependences of $\nu_{\min}$ and $\varepsilon_{\min}$ estimated in this way are presented in Figs. 5(f) and (g), respectively. Again, representations were chosen that should linearize the predicted MCT critical laws, Eqs. (5) and (6), as schematically indicated in Figs. 1(e) and (f). The lines in these figures are linear fits (ignoring the first data point, which is below $T_c$). They reasonably describe the data, thus indeed indicating the validity of the critical behavior of the minimum position predicted by MCT. One should note that the critical exponent $2a$ for $\nu_{\min}$, used in the ordinate of Fig. 5(f), is fixed by $\beta$ deduced from the right flanks of the loss peaks in Fig. 2(b) as discussed above. The $T_c$ values obtained from the fits, which again can be read off at the abscissa intercepts, are almost identical: 180 and 182 K for $\nu_{\min}$ and $\varepsilon_{\min}$, respectively. They are somewhat smaller than $T_c = 194$ K derived from $\nu_\alpha(T)$ of this solution [Fig. 5(a)]. However, considering the high uncertainties of the only roughly estimated MCT minimum curves in Fig. 2(b), we think the agreement is reasonable. Thus, we conclude that the indicated loss minima of the 14.8 mol% solution, together with its $\tau_\alpha(T)$, are well consistent with idealized MCT.

## C. Relaxation-times of water from different experimental methods

As mentioned above, in a recent work analyzing structural relaxation in micrometer-sized water droplets,[48] deviations of $\tau_\alpha(T)$ from the MCT critical law, Eq. (1), and a crossover to Arrhenius temperature-dependence below 233.2 K were reported. This was interpreted as evidence for an FST in water at this crossover temperature. Obviously, we cannot check this for the dielectric $\tau_\alpha(T)$ data on pure water (Fig. 2) whose lowest covered temperature is ~252 K. In contrast, for the solution with 14.8% LiCl, the $\tau_\alpha(T)$ trace nearly continuously extends far into the supercooled regime (Fig. 4). There, such a crossover from MCT behavior (solid line) to weaker temperature dependence indeed exists, read off in Fig. 4(a) to occur at roughly 210 K, somewhat above the divergence temperature $T_c = 194$ K. However, we want to point out that such deviations of relaxation-time data from the critical law of Eq. (1) occur for all materials analyzed with idealized MCT. As mentioned above, Eq. (1) is not expected to hold when approaching $T_c$ and there are numerous examples in literature documenting deviations of MCT fit and experimental data, analogous to those in Fig. 4.[40,41,89,106,107] Thus, following the logic in Ref. 48, an FST exists in all glass formers somewhat above $T_c$, simply marking the temperature below which idealized MCT no longer is applicable. Moreover, the MCT fit with Eq. (1), reported in that work, leads to a value for $\gamma$ of 0.82, far below the minimum value of ~1.76 allowed within MCT[2] [cf. section II and Fig. 1(c)]. It is also clearly at odds with $\gamma = 1.87$ of pure water, determined in the present work using dielectric spectroscopy, and with the critical exponents of water reported in literature, based on different experimental methods.[24,27,28,29,30,31] In the following, we reanalyze the $\tau_\alpha(T)$ data from Ref. 48 and compare them to the present dielectric data and those from other methods.

Figure 6(a) shows $\tau_\alpha(T)$ from Ref. 48. Analogous to the analysis in that work, the red line represents a fit of the data at $T > 233.2$ K using the MCT critical law, Eq. (1), with all parameters allowed to vary freely. While the fit is of good quality, we obtain $\gamma = 0.89$, far below the allowed minimum value of 1.76. The slight deviation from $\gamma = 0.82$ reported in Ref. 48 may be due to different fit algorithms or data-point weighting. The blue and green lines show fits with $\gamma$ fixed to its smallest possible value 1.76, performed for data points above 233.2 K or for all temperatures, respectively. Both are poor fits, and there is no clear evidence for an FST. Of course, it is possible to describe the small temperature range below 233.2 K by an Arrhenius law (i.e., a straight line in this representation) but this can be said for numerous regions of the experimental $\tau_\alpha(1/T)$ curve, e.g., for the leftmost six points ($T \geq 258$ K).

As the relaxation-time data of water from dielectric spectroscopy can be also well fitted by the VFT function [Eq. (7); dash-dotted line in Fig. 3(a)], in Fig. 6(b) we check this alternative approach for the data from Ref. 48. The solid line reveals that a very good fit by Eq. (7) is indeed possible. However, the obtained fit parameter $D = 0.311$ is unrealistically small. Via the relation $m \approx 16 + 590/D$, the fragility index $m$ can be calculated, which is the most common quantity for quantifying the degree of deviation from Arrhenius temperature dependence.[104] The obtained very small $D$ implies a fragility index $m \approx 1910$, far exceeding the accepted maximum value of $m$ which is about 170–200.[104,115] The dashed line in Fig. 6(b) demonstrates that fixing $D$ in Eq. (7) to 3.83 (i.e., $m \approx 170$), results in a very poor fit of $\tau_\alpha(T)$.

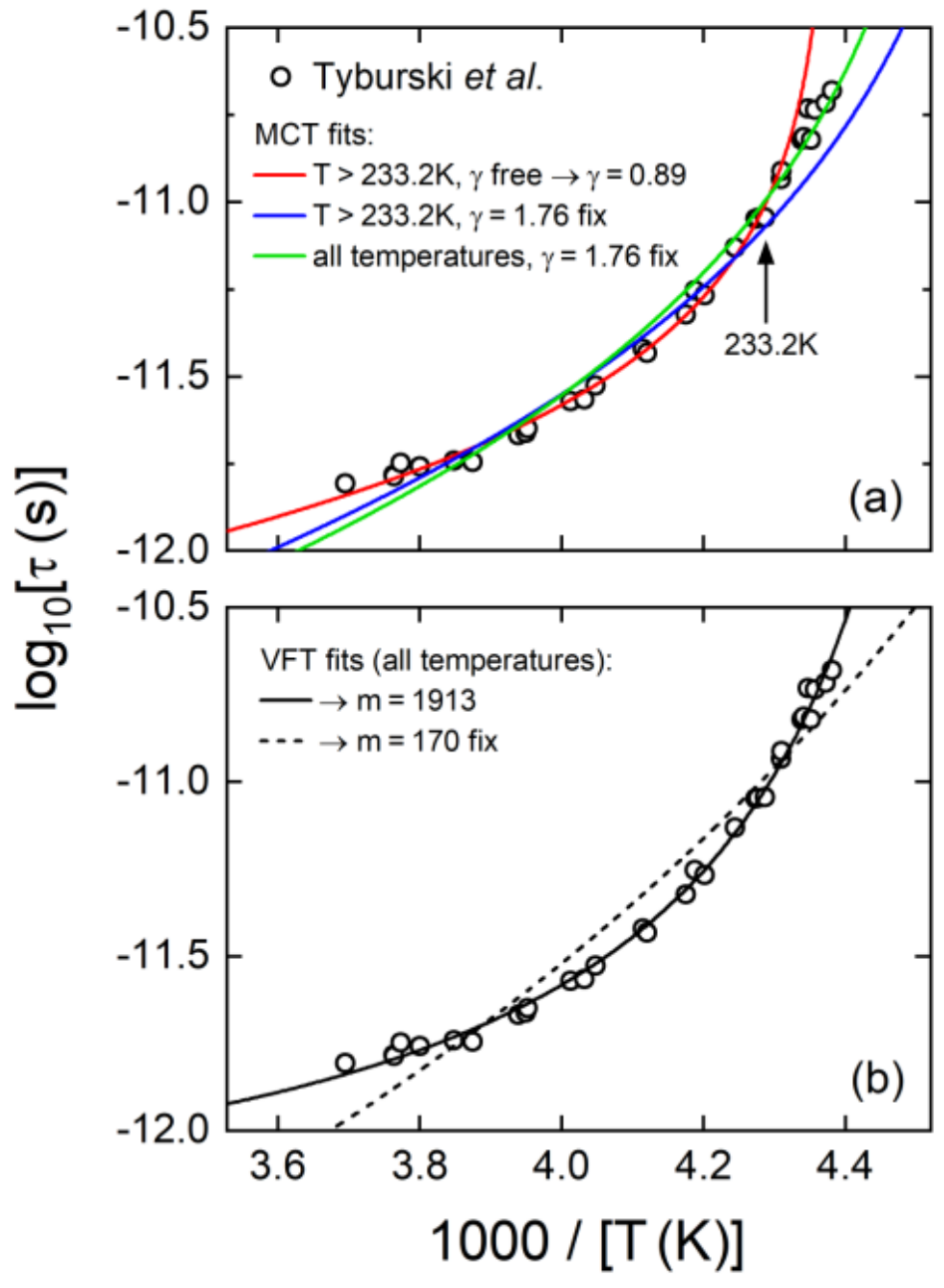

**FIG. 6.** Arrhenius plots of the structural relaxation times derived from ultrafast x-ray measurements on tiny water droplets.[48] The lines in (a) show three alternative fits with the MCT critical law, Eq. (1), as indicated in the legend. The lines in (b) are fits with the VFT function, Eq. (7). The solid line is a fit with all parameters free. For the dashed line, $D$ was fixed to 3.83, approximately corresponding to $m$ = 170.[104]

The question arises how these structural relaxation times from ultrafast measurements of water droplets[48] compare to the present dielectric results and to other relaxation-time data from literature. Figure 7 shows an Arrhenius plot of the dielectric $\tau_\alpha(T)$ data (plusses and crosses[21,92]) and of those from Ref. 48 (filled circles). Both differ by somewhat more than a decade. Generally, when checking the extensive literature on the relaxation dynamics in water studied by different methods, one finds that the absolute values of the reported relaxation times can vary considerably. To demonstrate this, a selection of such data is included in Fig. 7,[24,28,29,116,117,118,119,120,121,122] revealing variations up to a factor of 25. In some, but not all cases, the different logarithmic $\tau(1/T)$ traces are approximately parallel to each other, implying a fixed multiplicative factor connecting them. While similar deviations are known for other dipolar liquids[59,123,124] in some cases, e.g., glycerol, the relaxation times from different methods seem to agree quite well.[42,125] As revealed in Fig. 7, for water the shown NMR[29,120] and neutron-scattering data[117,121] are roughly located between those from dielectric spectroscopy and Ref. 48. In contrast, those from optical Kerr effect (OKE)[28] and mechanical modulus[116] are closer to the latter. The relaxation times from light-scattering[24,118,119,122] vary considerably, but at least those from Refs. 24 and 122 seem roughly compatible with those from Ref. 48. Interestingly, the flattening of $\tau_\alpha(T)$ from Ref. 48, seen at high temperatures ($T > 260$ K or $1000/T < 3.84$ K$^{-1}$), is not reproduced by any of the other data sets.

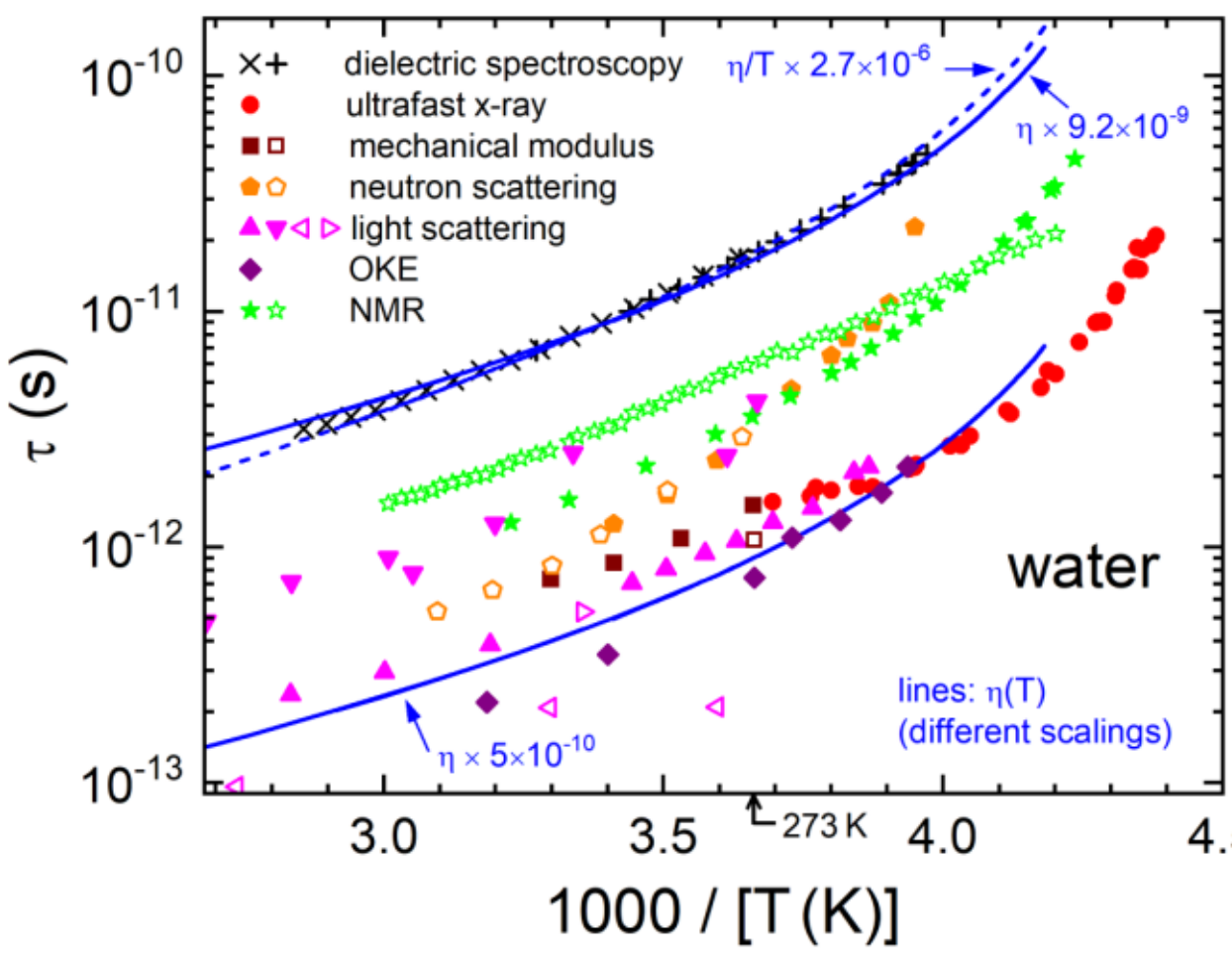

**FIG. 7.** Arrhenius plot of the relaxation times of water as detected by various experimental methods, indicated in the figure legend. The data were taken from Refs. 21 and 92 (dielectric spectroscopy; crosses and plusses, respectively), Ref. 48 (ultrafast x-ray; closed circles), Ref. 28 (OKE; closed diamonds), Refs. 24, 118, 119, and 122 (light scattering; up-, down, left, and right-pointing triangles, respectively), Refs. 117 and 121 (neutron scattering; closed and open pentagons, respectively), Ref. 116 (mechanical modulus; closed squares: volume, open square: shear), and Refs. 29 and 120 (NMR; closed and open stars, respectively). For comparison, the lines show the combined viscosity traces (in Pa s) from Refs. 96–98, scaled in three different ways as indicated in the figure (each of these three differently scaled traces in fact is composed of three separate lines from the three sources, however appearing as a single one due to their perfect agreement).

One should be aware that there are several rather trivial reasons that may explain at least part of the deviations revealed in Fig. 7. For example, relaxation-time values from susceptibility (dielectric, light and neutron scattering) and modulus spectra (mechanical) can differ considerably,[8,126] and differences in ranks of the correlation functions, associated with the methods, can influence $\tau_\alpha$.[122,127,128] In addition, relaxation times from neutron scattering can be difficult to compare to those from other techniques due to their momentum-transfer dependence,[129] and the identification of the $\alpha$ relaxation can be ambiguous if multiple processes are detected. However, more fundamental reasons for deviations in $\tau$ of water obtained from different techniques were proposed, too, especially for the finding of apparently slower $\alpha$-relaxation times by dielectric spectroscopy, compared to the other methods.[52,53] They essentially revive the old notion[49,50] that the dielectric response of water can be explained analogous to that of monohydroxy-alcohols,[130,131] where a dominant Debye relaxation, mostly undetected by other methods, is ascribed to the dynamics of clusters of hydrogen-bonded molecules.[132] In recent years, even ideas were promoted that, in practically all dipolar liquids, a dielectric-loss peak arises due to a similar mechanism, often termed "cross-correlations", which is often mistaken as the $\alpha$ peak. Within these scenarios, the "true" $\alpha$ relaxation, not affected by such inter-molecular correlations, only leads to a minor relaxation peak at higher frequencies.[125,133,134] Then the main dielectric loss peak of water would not represent the $\alpha$ relaxation. A detailed discussion of the above concepts is out of the scope of the present work. Nevertheless, we want to point out that, since long,[135] it is a quite common view that the $\alpha$ relaxation of liquids is not reflecting motions of independent molecules but instead is strongly influenced by "correlated" or "cooperative" molecular motions.[3,4,109]

## D. What is the "true" $\alpha$-relaxation time of water?

In light of the ideas mentioned in the preceding paragraph and considering the deviations revealed in Fig. 7, one may ask which method detects the "true" $\alpha$-relaxation time of water, to which MCT treating glassy freezing should be applicable. Due to the NML, this question cannot be solved by simply comparing the glass-transition temperatures, derived via $\tau_\alpha(T_g) \approx 100$ s, to those determined by calorimetric measurements as was done for some monohydroxy-alcohols.[136,137] Another option is checking for a match of the different $\tau(T)$ traces with the temperature dependence of the viscosity. On the latter, the original definition of the glass transition is based, namely the final solidification of a glass melt as observed by glass blowers since ancient times. Thus, a match of $\tau(T)$ and $\eta(T)$ would point to a close coupling of the relaxation dynamics characterized by $\tau$ to the glass transition. Such a comparison was also employed for the identification of the $\alpha$ relaxation in monohydroxy-alcohols.[130,138] The lines in Fig. 7 represent the viscosity literature-data of water,[96,97,98] scaled in three different ways: The upper solid line shows $\eta(1/T)$, simply multiplied by a fixed factor to match the dielectric $\tau_\alpha(1/T)$ (this is the same line as in the inset of Fig. 3). This leads to a reasonable agreement, implying $\tau_\alpha \propto \eta$, with very minor deviations due to a slightly smaller steepness of the shown $\eta$ trace. Plotting instead $\eta/T$, in accord with the sometimes-assumed relation $\tau_\alpha \propto \eta/T$,[138,139,140] somewhat overcompensates this small steepness difference (dashed line). Nevertheless, overall the temperature dependences of the dielectric relaxation time and viscosity can be considered as nearly identical.

The lower solid line in Fig. 7 indicates the same viscosity data, scaled by a different factor to come close to the $\tau(T)$ data from ultrafast x-ray measurements (closed circles).[48] The agreement is poor. Only if one ignores the mentioned shallow region at $T > 260$ K in the data from Ref. 48, a rough match with $\eta(T)$ may be stated. This is also the case for their compatibility with the OKE and part of the light scattering data. In principle, the Maxwell relation, $\tau_\alpha = \eta/G_\infty$, should provide absolute values for the scaling factor relating $\tau_\alpha$ and $\eta$. Here, $G_\infty$ is the limiting high-frequency shear modulus. Slie *et al*. determined $G_\infty$ of water at 273 K to be $\sim 1.68\times10^9$ Pa.[116] Its inverse ($\sim 6.0\times10^{-10}$ Pa$^{-1}$) is rather close to the multiplicative scaling factor of $5\times10^{-10}$ Pa$^{-1}$ used for the lower solid line in Fig. 7. Some caveat nevertheless applies: Slie's value for $G_\infty$ was deduced from an extrapolation of data of glycerol/water mixtures to zero glycerol content and, thus, has some uncertainty. Moreover, the Maxwell model leading to $\tau_\alpha = \eta/G_\infty$ is quite simplified and there are doubts about its applicability to water.[141,142]

Obviously, the identification of the "true" $\alpha$ relaxation depends on its definition. The scaling factor between $\eta$ and $\tau_\alpha$, derived from the Maxwell relation, indeed may raise some doubts whether water's dielectric relaxation process around 20 GHz reflects the $\alpha$ process. Nevertheless, it is by far the most dominant dynamics revealed in the dielectric spectra. As pointed out in Ref. 54, neutron-scattering results[129] imply that it "sets the longest relaxation timescale required for full equilibration of the liquid." In consideration of the fluctuation-dissipation theorem, the 20 GHz peak clearly reflects the most important dipolar dynamics within water. It is difficult to imagine molecular motions within water that do not also involve dipolar fluctuations. The latter are more or less directly detected by dielectric spectroscopy where ac fields are applied to the sample with frequencies that are (at the peak) identical to the fluctuation frequencies. In contrast, in light scattering, where the 20 GHz relaxation is essentially undetectable, the ac-field frequencies are many decades higher and molecular motions are more indirectly sensed via polarizability fluctuations.[127] In Ref. 88, analyzing the dielectric water spectra from Ref. 43 in terms of the coupling model and also considering results from other experimental methods, the main dielectric relaxation of water was assigned to the $\alpha$ process.

As shown above, the strongly dominating dipolar motion in water, marked by the 20 GHz dielectric loss peak, reveals non-Arrhenius and non-Debye behavior, both commonly considered as typical signatures of glassy freezing.[3,10] These features are also found for the main dipolar relaxations of numerous other glass forming liquids[17,42,104,106,143] to which MCT was successfully applied, leading to parameters that are well compatible with those derived from other experimental methods.[16,17,40,41,144] As demonstrated in the present work, in water the dominating dipolar relaxation mode is well consistent with the MCT predictions, too. It seems difficult to make all these finding compatible with recent ideas that the main dipolar relaxation in most dipolar liquids (including water) is governed by "cross-correlations" that are not related to the $\alpha$ relaxation, which is assumed to be significantly weaker and faster. Interestingly, based on MD simulations of aqueous solutions it was recently concluded that such cross-correlations become reduced upon addition of salt, "causing the total spectrum to increasingly resemble its self-part".[53] Nevertheless, in the present work we showed that the main dielectric relaxation processes of pure water and aqueous solutions can be coherently analyzed on the same footing, using MCT. This is corroborated by a smooth development of the MCT parameters with salt content [Figs. 5(d) and (e)].

Finally, we want to remark that, even when presuming that the 20 GHz mode is not related to the real $\alpha$ relaxation, it still reflects the dynamics of a by far dominant subsystem of dipolar entities in water that undergoes a glass transition. Our analysis demonstrates that this dynamics follows the predictions of the MCT for the $\alpha$ process. Other examples of subsystems with separate glass transition are the dipoles in plastic crystals and orientational glasses[145,146] or the ions in systems revealing decoupling of ionic and structural dynamics.[147,148]

## IV. SUMMARY AND CONCLUSIONS

In the present work, we have analyzed the most recent broadband dielectric-spectroscopy data on liquid water at temperatures above the NML and on LiCl solutions in terms of MCT. In particular, we have checked for the validity of the theoretical predictions concerning the spectral shapes of the $\alpha$ and fast $\beta$ relaxations and the critical laws that they should obey. Overall, all the available experimental results are well compatible with idealized MCT. This not only includes the critical temperature dependence of the $\alpha$-relaxation time, previously analyzed using various other spectroscopic methods, but also its connection to the often-neglected non-Debye shape of the major dielectric-loss peak of water. Moreover, we found that the available data on the fast $\beta$ process, detected in the GHz–THz range for a 14.8 mol% LiCl solution, are also well consistent with the theory. This includes its expected close interrelation to the $\alpha$ relaxation, showing up at much lower frequencies. The MCT parameters determined for pure water and the two investigated solutions reveal a smooth development in dependence of LiCl content, thus supporting the correctness of the critical temperature $T_c = 222$ K and system parameter $\lambda = 0.55$, derived for pure water.

In contrast to a recent work reporting an MCT analysis of the structural relaxation times of tiny water droplets,[48] the dielectric data do not provide evidence for a critical exponent $\gamma$ outside the allowed theoretical range or for an FST in bulk water at relatively high temperature around 230 K. $\tau(T)$ data from different experimental methods[24,27,28,29,30,31] also do not seem to support such behavior. A comparison of water's relaxation times as determined by various experimental methods reveals variations of up to 1.3 decades. Aside from trivial reasons for such deviations, obviously different methods are sensitive to different aspects of molecular motions in water. We have provided some arguments that the strongly dominating dielectric relaxation process of water around 20 GHz (at room temperature), as detected by dielectric spectroscopy, is closely related to the "true" $\alpha$ relaxation of water. However, in light of recent works,[52,53] it is clear that this is a disputed question. Anyway, based on the present results, we can at least clearly state that this main dipolar water relaxation well follows MCT's predictions made for the $\alpha$ relaxation. This agreement with theory is also valid for the solutions, detecting just the same main process. Notably, there the theoretical predictions are also consistent with the measured susceptibility in the THz region where the fast $\beta$ process is expected. Moreover, the obtained critical temperature of 222 K is in reasonable accord with $T_c$ values from other experimental methods,[24,27,28,29,30,31] assumed not to be hampered by the suggested cross-correlations. Indeed, according to MCT, the same $T_c$ should be obtained from all methods. Overall, MCT clearly applies for the dielectric water data. Whether they only reflect the dynamics of a glass-forming subsystem formed by the dominating dipoles or of the "true" structural $\alpha$ relaxation is an open question. Considering the present results, the latter seems more plausible to us.

Finally, one should note that there are more advanced versions of MCT, for example explicitly examining liquids containing asymmetric, rotating molecules and also treating the resulting susceptibilities from different methods.[149,150,151] It would be interesting to check whether such models may also be able to account for dielectric-spectroscopy data on water. Fabbian *et al.*[34] applied such a model to certain properties derived from MD simulations of supercooled water and found good agreement of its predictions with the MD results.

## AUTHOR DECLARATIONS

### Conflict of Interest

The authors have no conflicts to disclose.

## DATA AVAILABILITY

The data that support the findings of this study are available from the corresponding author upon reasonable request.